\documentclass[twocolumn,3p,final,times]{elsarticle}
\usepackage{graphicx,amsmath,amssymb}
\usepackage{hyperref}
\usepackage{breakurl}
\usepackage{color}
\usepackage{verbatim}
\usepackage{multirow}
\usepackage{tabularx}
\usepackage{slashed}
\usepackage{soul}
\usepackage[T1]{fontenc}
\usepackage{psfrag}

\newcommand{\eg}{\textit{e.g.}}  

\newcommand{\ie}{\textit{i.e.}}

\newcommand{\be}{\begin {align}}
\newcommand{\ee}{\end{align}}
\newcommand{\bi}{\begin{itemize}}
\newcommand{\ei}{\end{itemize}}
\newcommand{\bea}{\begin {eqnarray}}
\newcommand{\eea}{\end{eqnarray}}
\newcommand{\braket}[2]{\bra{#1}\,#2\rangle} 
\newcommand{\bra}[1]{\langle\,#1\,|}          
\newcommand{\ket}[1]{|\,#1\,\rangle}          
\newcommand{\ud}{\mathrm{d}}

\newcommand{\LCm}{{\scriptscriptstyle -}} 
\newcommand{\LCp}{{\scriptscriptstyle +}}

\newcommand{\LCperp}{{\scriptscriptstyle \perp}}

\definecolor{darkgreen}{rgb}{0.1,0.6,0}
\definecolor{gold}{rgb}{1,0.75,0}

\begin{document}

\title{Electron g-2 in Light-Front Quantization}

\author[IA]{Xingbo Zhao}
\ead{xbzhao@iastate.edu}
\author[HH]{Heli Honkanen}
\ead{hmh17@psu.edu}
\author[IA]{Pieter Maris}
\ead{pmaris@iastate.edu}
\author[IA]{James P. Vary}
\ead{jvary@iastate.edu}
\author[SJB]{Stanley J. Brodsky}
\ead{sjbth@slac.stanford.edu}

\address[IA]{Department of Physics and Astronomy, Iowa State University, Ames, Iowa 50011, USA}
\address[HH]{The Pennsylvania State University, 104 Davey Lab, University Park, PA 16802, USA}
\address[SJB]{SLAC National Accelerator Laboratory, Stanford University, Stanford, CA 94309, USA}

%
%
%
%

\date{\today}

\begin{abstract}
Basis Light-front Quantization has been proposed as a nonperturbative framework for solving quantum field theory. We apply this approach to Quantum Electrodynamics and explicitly solve for the light-front wave function of a physical electron. Based on the resulting light-front wave function, we evaluate the electron anomalous magnetic moment. Nonperturbative mass renormalization is performed. Upon extrapolation to the infinite basis limit our numerical results agree with the Schwinger result obtained in perturbation theory to an accuracy of 0.06\%.
\end{abstract}


\maketitle

\paragraph{Introduction}
\label{sec_intro}
Nonperturbative approaches in quantum field theory are needed for many applications. One important application is to study bound state problems in strongly interacting systems, \eg, solving for the hadron structure in Quantum Chromodynamics(QCD). The Basis Light-front Quantization (BLFQ) approach has recently been constructed~\cite{Vary:2009gt,Honkanen:2010rc} as a nonperturbative framework for quantum field theory in the Hamiltonian framework~\cite{Brodsky:1997de}. In previous work~\cite{Honkanen:2010rc} this method has been applied to QED for the anomalous magnetic moment of a physical electron in a cavity formed by an external transverse harmonic-oscillator potential, which acts as a regulator of QED dynamics. In addition, the extension to strong time-dependent external field applications (tBLFQ) has been developed and successfully applied~\cite{Zhao:2013cma,Zhao:2013jia}.

In  this work  we employ  BLFQ  to compute  the  electron anomalous  magnetic moment in vacuum,\footnote{In
addition, several improvements  and corrections are carried out  over Ref.~\cite{Honkanen:2010rc}: 1)
we correct the  operator used in~\cite{Honkanen:2010rc} for extracting the  anomalous magnetic moment
by adopting  the standard  Pauli form  factor operator~\cite{Brodsky:1980zm}; 
2)  we correct  a numerical  error in~\cite{Honkanen:2010rc},  which led  to an  overestimate of  the
interaction terms in QED Light-front Hamiltonian by a factor of 2;
3) we optimize the computational  efficiency on both the analytic and
coding level  and implement parallel  computing so  that the BLFQ  calculation can take  advantage of
currently  available supercomputers.  
} 
which was first calculated in perturbation theory in Ref.~\cite{Schwinger:1948iu}.
For  an alternative nonperturbative calculation  in light-front dynamics,
see Refs.~\cite{Chabysheva:2009ez,Chabysheva:2009vm}.


\paragraph{Light-front Hamiltonian in BLFQ basis}
\label{sec:review}
In light-front dynamics, physical processes are described in terms of light-front coordinates, which consist of light-front time
$x^+\equiv x^0+x^3$, the longitudinal direction $x^-$=$x^0-x^3$ as well as the transverse coordinates $x^\perp$=\{$x^1$,
$x^2$\}. We begin with the light-front QED Hamiltonian which can be derived from the ordinary Lagrangian through the standard
Legendre transform with the adoption of the light-cone gauge ($A^+$=0), where the photon field has physical polarization and positive metric. The resulting QED light-front Hamiltonian takes the following form,
\begin{align}
\label{eq:hami-full}
    P^- = &\int\!\ud^2x^\LCperp\ud x^\LCm \ \left[\frac{1}{2}\bar{{\Psi}} \gamma^\LCp \frac{m_e^2+(i\partial^\LCperp)^2}{i\partial^\LCp}\Psi + \frac{1}{2} { A}^j (i\partial^\LCperp)^2 { A}^j\right.\nonumber\\  &\left.+ e{j}^\mu {A}_\mu  + \frac{e^2}{2} { j}^\LCp \frac{1}{(i\partial^\LCp)^2}{ j}^\LCp + \frac{e^2}{2}\, \bar{\Psi} \gamma^\mu { A}_\mu \frac{\gamma^\LCp}{i\partial^\LCp} \gamma^\nu {A}_\nu \Psi\right]  \;,    	
\end{align}
where $\Psi$ and $A_\mu$ are the fermion and gauge boson fields, respectively. The first and second terms are their corresponding
kinetic energy terms, and the remaining three terms describe the interaction between the fermion and gauge boson
fields. Specifically, these are the {\it vertex interaction}, the {\it instantaneous-photon interaction} and the {\it
  instantaneous-fermion interaction} in order of appearance in Eq.~(\ref{eq:hami-full}). The $m_e$ and $e$ are the bare electron
mass and the bare electromagnetic coupling constant, respectively. In this work we only keep $\ket{e}$ and $\ket{e\gamma}$; \ie,
two Fock sectors, in our basis (see below). Consequently, the instantaneous-photon interaction does not contribute since it
involves Fock sectors with one more electron (or positron). Moreover, the instantaneous fermion interaction either contributes to
overall renormalization factors, which do not affect the intrinsic structure of the physical electron, or contains small-$x$
divergences which need to be cancelled by explicit fermion exchange contributions from higher Fock-sectors. Thus, we defer the
inclusion of the instantaneous interactions and adopt the following Hamiltonian for this work\footnote{The instantaneous
    fermion or photon interaction can also form a contact interaction in the single-electron sector, as mentioned in
    Ref.~\cite{Mustaki:1990im}. This type of interactions can, however, be absorbed into the fermion mass counterterm and does not need to be included explicitly in the Hamiltonian (see the {\it Renormalization} section below).},
\begin{align}
\label{eq:hami}
    P^- = &\int\!\ud^2x^\LCperp\ud x^\LCm \  \left[\frac{1}{2}\bar{{\Psi}} \gamma^\LCp \frac{m_e^2+(i\partial^\LCperp)^2}{i\partial^\LCp}\Psi + \frac{1}{2} { A}^j (i\partial^\LCperp)^2 { A}^j\right.\nonumber\\  &\left.+ e{j}^\mu {A}_\mu\right]  \;.
\end{align}

In the second step we construct the Fock-sector basis expansion. A physical electron, which is the focus of this work, receives contributions from the multiple Fock-sectors,
\begin{align}
\label{eq:Fock-sector-exp}
|e_\text{phys}\rangle=a|e\rangle+b|e\gamma\rangle+c|e\gamma\gamma\rangle+d|ee\bar{e}\rangle+\ldots,
\end{align}
and each Fock-sector itself consists of an infinite number of basis states. For the purpose of numerical calculations we adopt both a Fock-sector truncation and limits on the basis states in each sector. In this work we make the lowest nontrivial truncation by retaining the $\ket{e}$ and $\ket{e\gamma}$ Fock-sectors only.
This is sufficient for obtaining the (nonperturbative) electron wave function accurate to the leading order of the electromagnetic coupling constant $\alpha=\frac{e^2}{4\pi}$.


For each Fock-particle we employ a 2D-harmonic oscillator (HO) wave function, $\Phi_{nm}(p^\perp)$, to describe its transverse
degrees of freedom and a plane-wave, $e^{-i{ p^+ x^-/2}}$, to describe its longitudinal motion.  For each Fock-particle the
(transverse) 2D-HO wave function carries the radial quantum number $n$ and angular quantum number $m$, (as well as a
parameter $b$, setting the scale of the HO wave functions, \eg, in $\exp{[-(p^\perp)^2/(2b^2)]}$). The (longitudinal) plane-wave carries one quantum number, $k$=$p^+L/(2\pi)$, which is proportional to the longitudinal momentum $p^+$, and $L$ is the length of the longitudinal ``box'' in which we compactify the longitudinal degrees of freedom of the system. With the additional quantum number $\lambda$ for the helicity, 4 quantum numbers label each single particle state.

In the transverse directions we implement the ``$N_\text{max}$'' truncation in analogy with the 3D-HO truncation in nuclear many-body theory~\cite{Navratil:2000gs,Navratil:2000ww}. 
Define a sum, $N_\beta$, over the HO quantum numbers for all Fock particles, $i$, in a specific basis state (here we discuss only the quantum numbers of the transverse spatial motion), $\ket{\beta}$, according to,
\begin{align}
    N_\beta\equiv\sum_i 2n_i+| m_i |+1 \;.
\end{align}    
We truncate the basis states by eliminating states with $N_\beta$ larger than a chosen cutoff $N_\text{max}$. Increasing
$N_\text{max}$ not only enhances the resolution but also provides a higher ultraviolet cutoff and a lower infrared cutoff for the particles' transverse motion.

In the longitudinal direction we perform basis truncation by imposing (anti)periodic boundary condition for (fermions) bosons, such that the longitudinal momentum quantum number $k$ for each Fock particle can only take (half-)integers. Being a good quantum number for the QED Hamiltonian, the total longitudinal momentum summed over all Fock particles in a basis state, $P^+=\sum_i p^+_i\propto K$, acts as an additional cutoff. Larger $P^+(K)$ allows more possible partitions of longitudinal momentum among Fock particles in a basis state and thus leads to a higher resolution in the longitudinal degrees of freedom.

Therefore, in order to specify the truncated basis, we need the following information: 1) Fock-sectors included; 2) truncation
parameters, $N_\text{max}$ (transverse) and $K$ (longitudinal); 3) 2D-HO basis parameter $b$. The longitudinal period $L$ is not
needed due to the longitudinal boost-invariance of light-front dynamics: the light-front wave functions only depend on the
longitudinal momentum fraction $x_i=p^+_i/P^+=k_i/K$. In this work we retain only $\ket{e}$ and $\ket{e\gamma}$ Fock-sectors, and
compare numerical results evaluated in bases of different $N_\text{max}$ and $K$. Although in the $N_\text{max}\to\infty$ limit,
 results should be independent of $b$, we choose $b$=$M$=0.511\,MeV as the natural value for calculations at finite
 $N_\text{max}$. The (in-) dependence of $b$ in the numerical results for the electron anomalous magnetic moment $a_e$, around this value of $b$, will be checked in the {\it Anomalous Magnetic Moment} section below.

Next we express our field operators in the BLFQ basis, specifically for the fermion and gauge boson field,
\begin{align}
    \label{eq:field_op_momentum_basis_BLFQ}
    \Psi(\sf x) &= \sum_{\bar\beta}\frac{1}{\sqrt{2L}} \int\! \frac{\ud^2p^\perp}{(2\pi)^2} \big[b_{\bar\beta}\Phi_{nm}(p^\perp)u(\sf p,\lambda) e^{-i{\sf p\cdot x}} \nonumber \\
    &\qquad +d^\dagger_{\bar\beta} \Phi^*_{nm}(p^\perp)v(\sf p,\lambda) e^{i{\sf p\cdot x}} \big]  \;, \\
    A_\mu(\sf x)&=\sum_{\bar\beta}\frac{1}{\sqrt{2Lp^+}} \int\! \frac{\ud^2p^\perp}{(2\pi)^2} \big[ a_{\bar\beta}\Phi_{nm}(p^\perp)\epsilon_\mu(\sf p,\lambda)e^{-i{\sf p\cdot x}}  \nonumber \\
     &\qquad +a^\dagger_{\bar\beta}\Phi^*_{nm}(p^\perp)\epsilon^*_\mu(\sf p,\lambda)e^{i{\sf p\cdot x}}\big] \;,
\end{align}
where the $u$ and $v$ are the Dirac spinors for fermions and antifermions, respectively. The $\epsilon_\mu$ is the photon polarization vector. The ${\sf p\cdot x}=p^+x^-/2-p^\perp x^\perp$ is the inner product between the 3-momentum ${\sf p}=\{p^+,p^\perp\}$ and the coordinate ${\sf x}=\{x^-,x^\perp\}$. The $b^\dagger_{\bar\beta}$, $d^\dagger_{\bar\beta}$ and $a^\dagger_{\bar\beta}$ are creation operators for the fermion, antifermion and gauge boson fields, respectively, with quantum numbers $\bar\beta=\{k,n,m,\lambda\}$. They satisfy the (anti)commutation relations
\begin{align}
    \label{eq:commutation_relations}
	  \{ b_{\bar\beta},b^\dagger_{{\bar\beta}'} \} = \{ d_{\bar\beta},d^\dagger_{{\bar\beta}'} \} = [a_{\bar\beta},a^\dagger_{{{\bar\beta}'}}] = \delta_{\bar{\beta}\bar{\beta}'} \,.
\end{align}
Through Eqs.~(\ref{eq:hami}) and~(\ref{eq:commutation_relations}), we are able to write down the light-front QED Hamiltonian in the BLFQ basis. 
Since we are interested in the mass eigenspectrum contributed by the intrinsic rather than center-of-mass motion, we add an appropriate Lagrange multiplier term to the input light-front QED Hamiltonian.  This has the effect of shifting the states with excited center-of-mass motion to high mass and the low-lying spectrum comprises states with lowest center-of-mass motion, following the techniques of nuclear many-body theory. The resulting low-lying states can be written as a simple product of internal and center-of-mass motion ~\cite{Navratil:2000gs,Navratil:2000ww} (see~\cite{Zhao:2013xx} for more details).

Upon diagonalization of the resulting sparse Hamiltonian matrix, one obtains its eigenvalue spectrum and corresponding eigenvectors. In this work, the ground state of the Hamiltonian, with net fermion number being one ($n_f{=}1$), is identified as the physical electron. Its eigenvalue, $P^-_e$, gives the electron mass according to $M^2\equiv P^-_eP^+_e-P^2_\perp$, where $P_\perp$ is the total transverse momentum operator.

\paragraph{Renormalization}
\label{sec:renorm}
Before we are ready to obtain the electron wave function, one more technical detail needs to be worked out:
renormalization. In BLFQ, a nonperturbative approach, the renormalization procedures are different from
  those adopted in perturbation theory~\cite{Brodsky:1973kb}. 

In quantum field theory, the values for parameters in the Hamiltonian, the bare electron mass $m_e$ and the bare coupling
constant $e$, are regulator (cutoff) dependent. Through renormalization, one establishes the exact connection between these
parameters and the theory's regulators. Since we omit Fock-sectors containing electron-positron pairs in our bases, bare photons
are not able to fluctuate into electron-positron pairs and thus modify the physical charge of the
electron\footnote{Strictly speaking, this statement is true only if the Ward-identity holds. Here, the Ward-identity is
broken by Fock-sector truncations. In the literature there exist methods~\cite{Karmanov:2008br,Karmanov:2012aj} where the electron
charge renormalization is invoked to rectify the artifacts caused by the loss of the Ward-identity, while other approaches,
\eg,~\cite{Brodsky:2004cx}, elect other methods to achieve the same goal. In this paper, we choose to use the ``rescaling'' of the
physical electron wavefunction (see below) to remedy the artifacts caused by the loss of the Ward-identity.}. 
In this work we need only the electron mass renormalization.

Guided by a sector-dependent renormalization approach~\cite{Karmanov:2008br,Karmanov:2012aj}, our procedure for the electron mass renormalization is as follows: we numerically diagonalize the Hamiltonian matrix in an iterative scheme where we adjust the input bare electron mass in the Hamiltonian in the single electron sector only, until the resulting mass for the ground (physical electron) state matches the physical electron mass of $M$=0.511\,MeV. The idea behind this procedure is the following: the mass counterterm, the difference between the physical electron mass and the bare mass, compensates for the mass correction due to the quantum fluctuations to higher Fock sectors. The basis states in the electron-photon sector, the highest Fock sector in our current truncation scheme, generate the conventional one-loop self-energy interactions. No further quantum fluctuations are introduced. Thus for basis states in the electron-photon sector the bare electron mass remains the same as the physical value. On the other hand, the basis states in the single electron sector couple to those in the electron-photon sector and receive the self-energy correction. Therefore for these states, we need a mass counter-term which we introduce via our iterative diagonalization scheme.

\paragraph{Anomalous Magnetic Moment}
\label{sec:anoma}
Our calculated spectrum includes both the physical electron state and electron-photon scattering states. The ground state is identified as the physical electron ($\ket{e_\text{phys}}$), and its eigenvalue has been renormalized to the mass of a physical electron. The associated eigenvector (wave function) encodes all the information of  intrinsic structure of the physical electron and can be employed to evaluate observables.

We focus on one specific observable: the electron anomalous magnetic moment, $a_e$, which measures the deviation of the electron spin gyromagnetic ratio from the ``normal'' value, 2, namely, $a_e\equiv\frac{g_s-2}{2}$. The electron spin gyromagnetic ratio $g_s$ is the ratio between the electron's magnetic moment, $\mu$, and the product of the electron's spin, $s$=1/2, with the Bohr magneton, $e/(2M)$,
\begin{align}
    \mu=g_s\frac{e}{2M}s\; .
\end{align}
The finite electron anomalous magnetic moment $a_e$ reflects a nontrivial internal structure of the electron in QED: it originates from the relative motion between the constituent electron and the constituent photon (as well as higher Fock components in principle) inside a physical electron. It was first calculated by Schwinger in leading-order perturbation theory~\cite{Schwinger:1948iu} with the result $a_e=\frac{\alpha}{2\pi}$.

In QED, the $a_e$ is defined by the Pauli form factor $F_2(q^2)$
at the zero momentum transfer limit $q^2\to0$,
\begin{align}
    \label{eq:anoma}
        a_e\equiv F_2(q^2\to0)\; ,
\end{align}
where $q^2=q^\mu q_\mu$ and $q^\mu$ is the 4-momentum transferred from a probe photon to the electron. We adopt the Drell-Yan-West frame~\cite{Brodsky:1980zm} where the incident electron is directed along the 3-direction with 4-momentum $p^\mu=(p^+,\frac{M^2}{p^+},\vec{0}_\perp)$ and the probe photon's momentum is in the transverse directions with $q^\mu=(0,\frac{2q\cdot p}{p^+},\vec{q}_\perp)$, where $2q\cdot p=-q^2=q^2_\perp$.
In this frame the Pauli form factor can be evaluated as,
\begin{align}
    \label{eq:f2_eval}
    -\frac{q_1-iq_2}{2M}F_2(q^2)=\bra{e^\uparrow_\text{phys}(\vec{q}_\perp)}J^+(0)\ket{e^\downarrow_\text{phys}(\vec{0}_\perp)},
\end{align}
where $\vec{q}_\perp{=}(q_1,q_2)$ and $J^+(0){=}\bar{\Psi}(0)\gamma^+\Psi(0)$ is the electric charge density operator at $x^\mu$=0. The
$\ket{e^{\uparrow(\downarrow)}_\text{phys}(\vec{q}_\perp)}$ denotes the physical electron state with helicity (anti) parallel to the
longitudinal momentum ($p^+$) direction and (average) transverse center-of-mass momentum of $\vec{q}_\perp$. 
The helicity-flip state and the states with nonzero (average) transverse momentum can be inferred from $\ket{e^{\uparrow}_\text{phys}(\vec{0}_\perp)}$ by exploiting the transverse parity symmetry~\cite{Brodsky:2006ez,Zhao:2013xx} and the boost invariance properties of light-front dynamics. 
 


\begin{figure}[!t]
\psfrag{amm}[b]{$\sqrt{a_e/e^2}$ from Eq.~(\ref{eq:f2_eval})}
\psfrag{1/Z2}[b]{$1/Z_2$}
\psfrag{alpha1137}{$\alpha=\frac{1}{137.036}$}
\psfrag{alpha1pi}{$\alpha=\frac{1}{\pi}$}
\psfrag{fitalpha1137}{$0.87+0.044\ln{N_\text{max}}$}
\psfrag{fitalpha1pi}{$-4.0+1.78\ln{N_\text{max}}$}
\psfrag{NmaxK}{\hspace{-6mm}$N_\text{max}=K-1/2$}
\includegraphics[width=0.48\textwidth]{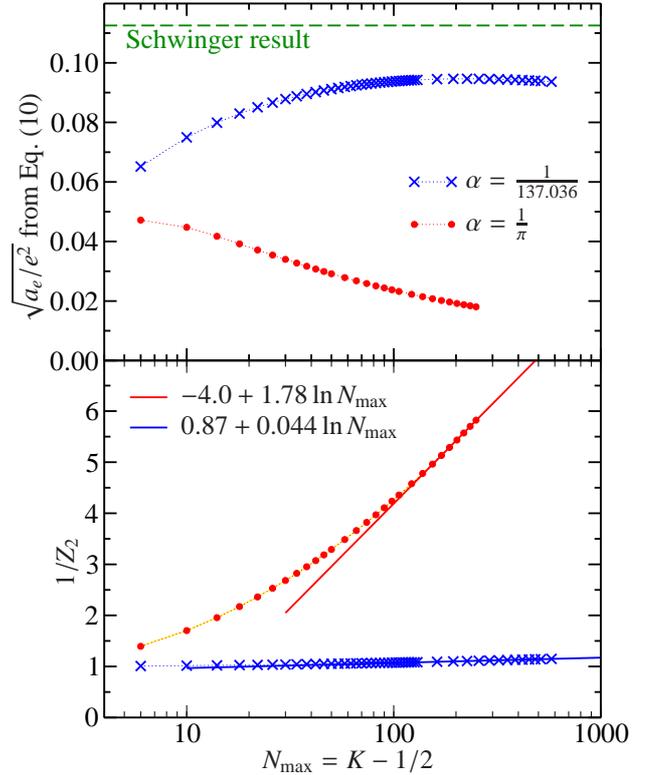}
\caption{(Color online) Upper panel: the (square root of) electron anomalous magnetic moment $a_e$ (normalized to electron charge
  $e$) from Eq.~(\ref{eq:f2_eval}) as a function of basis truncation parameter $N_\text{max}{=}K{-}1/2$. All the data points are
  at odd $N_\text{max}/2$. The dashed line indicates the Schwinger result = 0.1125395. Lower panel: the (inverse) electron
  wave-function renormalization factor $Z_2$ as a function of basis truncation parameter $N_\text{max}{=}K{-}1/2$. The
  two-parameter fits, indicated in the legend, are based on data points with $N_\text{max}{=}K{-}1/2>150$ only. We use $b$=$M$ for the results in both panels.}
\label{fig:anoma_Z2}
\end{figure}

In BLFQ, we work with finite dimensional basis spaces. In order to obtain  the electron anomalous magnetic moment
$a_e$ in the limit of the infinite basis size, we  first calculate $a_e$ as a function of the truncation parameters, which also act as regulators, and then perform  extrapolations. The upper panel of Fig.~\ref{fig:anoma_Z2} displays the anomalous magnetic moment evaluated from Eq.~(\ref{eq:f2_eval}) at various selected basis sizes as discrete points and at both the physical electromagnetic coupling constant $\alpha=\frac{1}{137.036}$ and an artificially enlarged $\alpha=\frac{1}{\pi}$. We elect to relate the two basis space cutoffs and adopt $N_\text{max}$ = $K{-}1/2$ for simplicity and convenience. The horizontal axis in Fig.~\ref{fig:anoma_Z2} is the $N_\text{max}=K{-}1/2$ of the basis.
Results in bases at $N_\text{max}{=}K{-}1/2$ larger than 50 are evaluated using Hopper, a Cray XE6 supercomputer,
and Edison, a Cray XC30 supercomputer, at the National Energy Research Scientific Computing Center (NERSC). Numerical diagonalization of the Hamiltonian matrix is performed by ARPACK software~\cite{Lehoucq97arpackusers}.
The maximum basis dimensionality achieved so far is 28,027,289,920 at $N_\text{max}=K{-}1/2$=640. 

The results in the upper panel of Fig.~\ref{fig:anoma_Z2} suggest that the anomalous magnetic moment directly evaluated from Eq.~(\ref{eq:f2_eval}) tends to zero with increasing basis space cutoff.
The origin of this behavior is that the current Fock space truncation violates the condition $Z_1$ = $Z_2$~\cite{Brodsky:2004cx}, which would be a consequence of the Ward identity. Here $Z_1$ is the renormalization factor for the vertex coupling the $\ket{e}$ and $\ket{e\gamma}$ sectors which remains unity
in the infinite basis limit with our Fock space truncation.  Now, $Z_2$ is the electron wave-function renormalization which, in light-front dynamics, can be interpreted as the probability of finding a constituent electron out of a physical electron:
\begin{align}
    \label{eq:Z_2}
    Z_2 = \sum_{\ket{e}} |\braket{e}{e_\text{phys}}|^2,
\end{align}
where the summation runs over all the basis states in the $\ket{e}$ sector. In our Fock space truncation, $Z_2$ receives a contribution from the quantum fluctuation between the $\ket{e}$ and $\ket{e\gamma}$ sectors and consequently goes to zero in the infinite basis limit. Our numerical data suggest a logarithmic divergence in $1/Z_2$ as a function of the truncation parameters (regulators), see the lower panel of Fig.~\ref{fig:anoma_Z2}.

\begin{figure}[!t]
%
%
\begin{minipage}[c]{0.482\textwidth}
\includegraphics[width=1.0\textwidth]{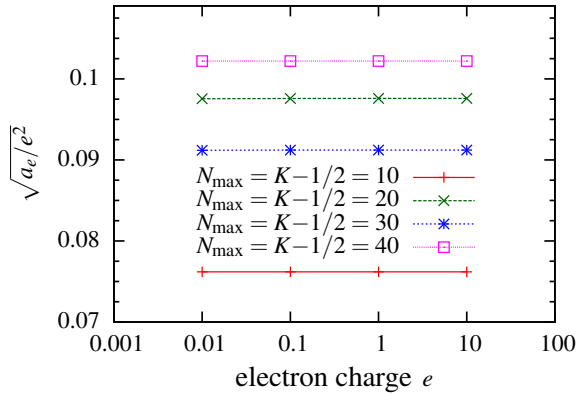}
\end{minipage}
\caption{(Color online) The (square root of) electron anomalous magnetic moment $a_e$ (normalized to electron charge $e$) as a
  function electron charge $e$ at selected $N_\text{max}=K{-}1/2$. At each $N_\text{max}=K{-}1/2$, the variation is in the fourth
  significant digit and not visible on the figure. We use $b$=$M$.}
\label{fig:g-dep}
\end{figure}

We next note that, due to our current Fock space truncation, $Z_1$  does not obtain the corresponding quantum fluctuation that would involve the $\ket{e\gamma\gamma}$ sector.  
Hence, it seems reasonable to associate the origin of the vanishing (naive) anomalous magnetic moment from Eq.~(\ref{eq:f2_eval}) with that of the vanishing renormalization constant $Z_2$.
We therefore propose the following procedure to obtain the rescaled (``re'') Pauli form factor,
\begin{align}
    \label{eq:anoma_renorm}
    a_e=F^{\text{re}}_2(0)=\frac{F_2(0)}
    {Z_2}.
\end{align}
After rescaling the Pauli form factor the (rescaled) anomalous magnetic moment becomes independent of the coupling constant $\alpha$ (at fixed $N_\text{max} = K -1/2$), as can be seen in Fig.~\ref{fig:g-dep}, even though the naive results for the anomalous magnetic moment depend strongly on $\alpha$ (see Fig.~\ref{fig:anoma_Z2}).  Furthermore, our results for the rescaled anomalous magnetic moment seem to increase monotonically with increasing $N_\text{max} = K - 1/2$, and approach the Schwinger result from below, independent of $\alpha$. 

\begin{figure}[!t]
\includegraphics[width=0.48\textwidth]{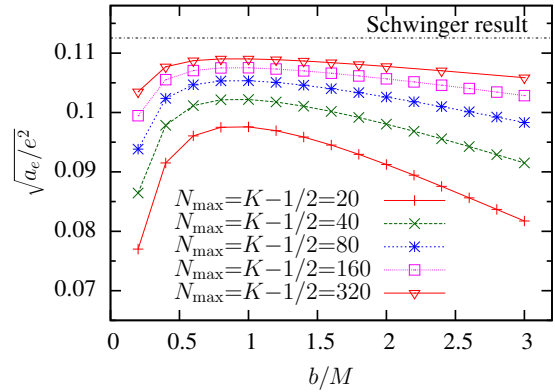}
\caption{(Color online) The (square root of) electron anomalous magnetic moment $a_e$ (normalized to electron charge $e$) as a
  function of the 2D-HO basis parameter $b$ (in the unit of electron mass $M$). The horizontal dot-dashed line indicates the
  Schwinger result = 0.1125395. At $N_\text{max}=K-1/2=320$, we opted to calculate fewer points to save on the computational resources.}
\label{fig:anoma_b}
\end{figure}

Next, we check the dependence of the rescaled $a_e$ on the
2D-HO basis parameter $b$ in Fig.~\ref{fig:anoma_b}. As we increase
$N_\text{max}=K-1/2$, the results show improved independence of $b$  
over an increasingly large interval centered around $b=M$. This improving
independence of $b$ is a signal for convergence with increasing
basis dimension.

\begin{figure}[!t]
\includegraphics[width=0.48\textwidth]{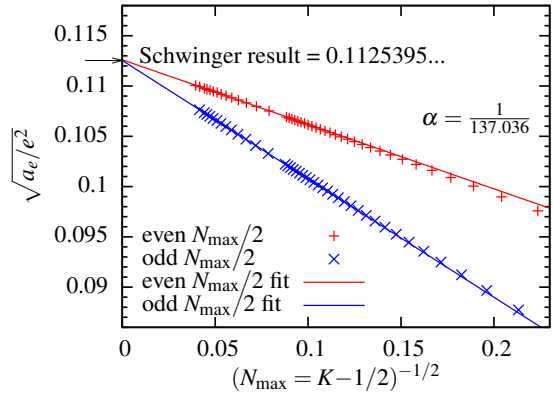}
\caption{(Color online) The (square root of) electron anomalous magnetic moment $a_e$ (normalized to electron charge $e$) as a
  function of basis truncation parameter $N_\text{max}{=}K{-}1/2$. We use $\alpha=\frac{1}{137.036}$ and $b$=$M$. Linear extrapolations use data points with $N_\text{max}{=}K{-}1/2$>150. The arrow on the left y-axis indicates the Schwinger result = 0.1125395. }
\label{fig:anoma_vs_nmax}
\end{figure}

In order to test the precision of BLFQ, we extrapolate the (rescaled) anomalous magnetic moment to the infinite $N_\text{max}=K-1/2$ limit in Fig.~\ref{fig:anoma_vs_nmax}.
Here, the BLFQ results fall into two groups with even and odd $N_\text{max}/2$, respectively. This odd-even effect originates from the oscillatory behavior of the (2D-HO) basis function in the transverse plane. 
In Fig.~\ref{fig:anoma_vs_nmax}, we apply linear extrapolation in $1/\sqrt{N_\text{max}=K-1/2}$ to data points with $N_\text{max}{=}K$>150 for the even (odd) $N_\text{max}$/2 group individually. The extrapolated $a_e$ at infinite $N_\text{max}=K{-}1/2$ limit, is 0.112610 (0.112541), agreeing well with the Schwinger result with a relative deviation of $+$0.063\% ($+$0.001\%), for the even (odd) $N_\text{max}$/2 group, respectively. 

Our finding confirms the analytic result found in Refs.~\cite{Karmanov:2008br,Karmanov:2012aj}, where a nonperturbative light-front wave equation approach is adopted and it was found that in bases truncated to $\ket{e}$ and $\ket{e\gamma}$ two sectors, the nonperturbative results agree exactly with the Schwinger result.
This can be understood as follows: in the Fock-sector truncation allowing only for the quantum fluctuation into $\ket{e\gamma}$ sector, the resulting nonperturbative light-front wavefunction encodes the identical information on the structure of the physical electron, compared to that from leading-order perturbation theory. The higher-order contributions only contribute to the electron wave-function renormalization factor, $Z_2$.
Once we include higher Fock sectors in our BLFQ calculations we expect to see deviations between the nonperturbative BLFQ results and (higher-order) perturbation theory.
\paragraph{Conclusion and Outlook}
\label{sec:end}

In this work we demonstrated the workflow of applying BLFQ to evaluate observables in the vacuum. Specifically, as a test problem, we applied this approach to QED and study the physical electron in bases truncated to $\ket{e}$ and $\ket{e\gamma}$ Fock-sectors. We performed the electron mass renormalization following a sector-dependent scheme~\cite{Karmanov:2008br,Karmanov:2012aj}.
We found that the resulting (naive) electron anomalous magnetic moment in this truncated basis approaches zero upon extrapolation
to the infinite basis limit, independent of the coupling constant. However, by rescaling the anomalous magnetic moment with the
inverse of the electron wave-function renormalization factor ($Z_2$), we recover the Schwinger result to high precision (less than
0.1\% deviation), confirming the results found in another nonperturbative approach based on light-front wave equation
formalism~\cite{Karmanov:2008br,Karmanov:2012aj}.

Renormalization in the nonperturbative Hamiltonian formalism of quantum field theory is a long-standing problem, mainly due to the fact that a Fock-sector truncation breaks the Ward-identity (see, \eg, Ref.~\cite{Brodsky:2004cx} for a detailed discussion). 
We introduce a ``rescaling'' procedure as an initial attempt to address this problem and
we verify this procedure by a high precision numerical calculation of the electron's 
anomalous magnetic moment. Although we are not able to provide a full theoretical justification for
this procedure at this time, we believe that the ultimate validity or our approach will become more clear when additional problems of similar nature are solved in the future.



On the computational aspects, we find the BLFQ approach may be parallelized following recent advances in computational low-energy nuclear physics. For fixed basis sizes, the (inverse) computational time (``speedup'' factor) scales almost linearly with the number of processors. It is thus conceivable that this method will greatly benefit from anticipated advances in supercomputer technology.

Since the electron light-front wave function encodes all the information on the electron structure, it can be employed to evaluate other observables which ``measure'' the electron structure in QED, such as the electromagnetic form factors, the generalized parton distribution functions (GPDs)~\cite{Chakrabarti:2014cwa}, etc. Also, we have initiated applications of this method to other systems, such as positronium, for which we add a positron into the current single electron system. Indeed, initial positronium calculations are already underway~\cite{Maris:2013qma,Wiecki:2013cba,Li:2013cga}.
In addition to these ``stationary'' problems, electron light-front solutions also find applications in the recently developed time-dependent Basis Light-front Quantization (tBLFQ) approach~\cite{Zhao:2013cma}, where the single electron states as well as the electron-photon scattering states are employed to solve the photon-emission problem in a strong and time-dependent laser field. Ultimately, our goal is to apply this method to QCD and solve for the hadron spectrum and structure.

We acknowledge valuable discussions with K. Tuchin, P. Hoyer, P. Wiecki and Y. Li. This work was supported in part by the U.S. Department of Energy under Grant Nos. DE-FG02-87ER40371, DESC0008485 (SciDAC-3/NUCLEI) , DE-FG02-93ER40771 and DE-AC02-76SF00515 and by the National Science Foundation under Grant No. PHY-0904782.
A portion of the computational resources were provided by the National Energy Research Scientific
Computing Center (NERSC), which is supported by the Office of Science of the U.S. Department of Energy under Contract
No. DE-AC02-05CH11231. 


\begin{thebibliography}{99}

\bibitem{Vary:2009gt}
  J.~P.~Vary {\it et al.},
  Phys.\ Rev.\  C {\bf 81},(2010) 035205. 

\bibitem{Honkanen:2010rc}
  H.~Honkanen, P.~Maris, J.~P.~Vary and S.~J.~Brodsky,
  Phys.\ Rev.\ Lett.\  {\bf 106}, 061603 (2011).
  
\bibitem{Brodsky:1997de}
  S.~J.~Brodsky, H.~-C.~Pauli and S.~S.~Pinsky,
  Phys.\ Rept.\  {\bf 301} (1998) 299.
  
\bibitem{Zhao:2013cma} 
  X.~Zhao, A.~Ilderton, P.~Maris and J.~P.~Vary,
  Phys.\  Rev.\  D {\bf 88}, 065014 (2013).
  
\bibitem{Zhao:2013jia}
 X.~Zhao, A.~Ilderton, P.~Maris and J.~P.~Vary,  
 Phys.\ Lett.\ B {\bf 726} (2013) 856.
 

  \bibitem{Brodsky:1980zm} 
  S.~J.~Brodsky and S.~D.~Drell,
  Phys.\ Rev.\ D {\bf 22}, 2236 (1980).

\bibitem{Schwinger:1948iu}
  J.~S.~Schwinger,
  Phys.\ Rev.\  {\bf 73} (1948) 416.
  
%
\bibitem{Chabysheva:2009ez} 
  S.~S.~Chabysheva and J.~R.~Hiller,
Annals Phys.\  {\bf 325}, 2435 (2010).


\bibitem{Chabysheva:2009vm} 
  S.~S.~Chabysheva and J.~R.~Hiller,
Phys.\ Rev.\ D {\bf 81}, 074030 (2010).


  
  


\bibitem{Mustaki:1990im} 
  D.~Mustaki, S.~Pinsky, J.~Shigemitsu and K.~Wilson,
  Phys.\ Rev.\ D {\bf 43}, 3411 (1991).

  

\bibitem{Navratil:2000gs}
  P.~Navratil, J.~P.~Vary and B.~R.~Barrett,
  Phys.\ Rev.\ C {\bf 62} (2000) 054311.

\bibitem{Navratil:2000ww} 
  P.~Navratil, J.~P.~Vary and B.~R.~Barrett,
  Phys.\ Rev.\ Lett.\  {\bf 84}, 5728 (2000).
  

\bibitem{Zhao:2013xx}
  X.~Zhao, H.~Honkanen, P.~Maris, J.~P.~Vary and S.~J.~Brodsky,
  {\it in preparation}.   
  

\bibitem{Brodsky:1973kb} 
 S.~J.~Brodsky, R.~Roskies and R.~Suaya,
 Phys.\ Rev.\ D {\bf 8}, 4574 (1973).


\bibitem{Karmanov:2008br} 
  V.~A.~Karmanov, J.~-F.~Mathiot and A.~V.~Smirnov,
Phys.\ Rev.\ D {\bf 77}, 085028 (2008).


\bibitem{Karmanov:2012aj} 
  V.~A.~Karmanov, J.~-F.~Mathiot and A.~V.~Smirnov,
	Phys.\ Rev.\ D {\bf 86}, 085006 (2012).


\bibitem{Brodsky:2004cx} 
  S.~J.~Brodsky, V.~A.~Franke, J.~R.~Hiller, G.~McCartor, S.~A.~Paston and E.~V.~Prokhvatilov,
  Nucl.\ Phys.\ B {\bf 703}, 333 (2004).
    
\bibitem{Brodsky:2006ez} 
  S.~J.~Brodsky, S.~Gardner and D.~S.~Hwang,
  Phys.\ Rev.\ D {\bf 73}, 036007 (2006).


  
\bibitem{Lehoucq97arpackusers} 
  R.~B.~Lehoucq and D.~C.~Sorensen and C.~Yang,
  {\it ARPACK Users Guide: Solution of Large Scale Eigenvalue Problems by Implicitly Restarted Arnoldi Methods}, (1997).  
  

  

\bibitem{Chakrabarti:2014cwa}
  D.~Chakrabarti, X.~Zhao, H.~Honkanen, R.~Manohar, P.~Maris and J.~P.~Vary,
  Phys.\ Rev.\ D {\bf 89} (2014) 116004.
  
\bibitem{Maris:2013qma} 
  P.~Maris, P.~Wiecki, Y.~Li, X.~Zhao and J.~P.~Vary,
  Acta Phys.\ Polon.\ Supp.\  {\bf 6}, 321 (2013).
  
\bibitem{Wiecki:2013cba} 
  P.~W.~Wiecki, Y.~Li, X.~Zhao, P.~Maris and J.~P.~Vary,
  in the Proc. of Int. Conf. `Nuclear Theory in the Supercomputing
  Era -- 2013' (NTSE-2013), Ames, IA, USA, May 13-17, 2013. Eds. A.M. Shirokov
  and A.I. Mazur. Pacific National University, Khabarovsk, Russia (2014) 146, ISBN 978-5-7389-1384-6,
  http://www.ntse-2013.khb.ru/Proc/Wiecki.pdf.

\bibitem{Li:2013cga} 
  Y.~Li, P.~W.~Wiecki, X.~Zhao, P.~Maris and J.~P.~Vary, 
  in the Proc. of Int. Conf. `Nuclear Theory in the Supercomputing
  Era -- 2013' (NTSE-2013), Ames, IA, USA, May 13-17, 2013. Eds. A.M. Shirokov
  and A.I. Mazur. Pacific National University, Khabarovsk, Russia (2014) 136, ISBN 978-5-7389-1384-6,
  http://www.ntse-2013.khb.ru/Proc/YLi.pdf.

%
%
%
%
%
%
%
%
  
 
  
%
  
  
  
  


%
  

%
%
  

  
  
  
  
  


  
  
  





  




%
  






%
%
%
%
%
  


%



%
%
%




%







 
%
%

  
  
%

\end{thebibliography}
\end{document}